\documentstyle[12pt,a41,amssymb,epsfig]{article}
\begin{document}
\oddsidemargin -0.65cm
\evensidemargin -0.65cm
\thispagestyle{empty}
\begin{flushright}
\large
CERN-TH/97-310 \\
DTP/97/98 \\
UCL/HEP 97-08 \\
November 1997
\end{flushright}
\vspace{1.cm}

\begin{center}
\LARGE
{\bf Photoproduction Processes }

\vspace*{0.5cm}
{\bf in Polarized $ep$ - Collisions at HERA }

\vspace{1.8cm}
{\Large J.M.\ Butterworth$^a$, N.\ Goodman$^a$, M.\ Stratmann$^b$, 
W.\ Vogelsang$^c$ }

\vspace*{1.5cm}
\large
{\it $^a$Department of Physics and Astronomy, University College London,}

\vspace*{0.1cm}
{\it Gower Street, London WC1E 6BT, England}

\vspace*{0.1cm}
{\it $^b$Department of Physics, University of Durham, Durham DH1 3LE, England}

\vspace*{0.1cm}
{\it $^c$Theoretical Physics Division, CERN, CH-1211 Geneva 23, Switzerland}

\vspace{2.2cm}
{\bf Abstract} 
\end{center}

\noindent
\normalsize
We study various conceivable photoproduction reactions in a polarized
$ep$ collider mode of HERA with respect to their sensitivity to the
proton's polarized gluon distribution. A special emphasis is put on
the `resolved' part of the cross sections which in principle opens the
possibility to determine for the first time also the completely
unknown parton content of longitudinally polarized photons. In the
very promising case of dijet production we also investigate the impact
of parton showering, hadronization and jet finding on the parton level
results.

\vspace{4.0cm}
\noindent
{\sf Contribution to the proceedings of the 1997 workshop on 
`Physics with Polarized Protons at HERA', DESY-Hamburg and
DESY-Zeuthen, March-September 1997.} 
\vfill

\setcounter{page}{0}
\newpage
\begin{center}
{\LARGE \bf Photoproduction Processes }

\vspace*{0.3cm}
{\LARGE \bf in Polarized $ep$ - Collisions at HERA }

\vspace{1cm}
{\Large J.M.\ Butterworth$^a$, N.\ Goodman$^a$, M.\ Stratmann$^b$, 
W.\ Vogelsang$^c$ }

\vspace*{1cm}
{\it $^a$Department of Physics and Astronomy, University College London,\\
Gower Street, London WC1E 6BT, England \\
$^b$Department of Physics, University of Durham, Durham DH1 3LE, England\\
$^c$Theoretical Physics Division, CERN, CH-1211 Geneva 23, Switzerland}

\vspace*{0.7cm}

\end{center}

\begin{abstract}
We study various conceivable photoproduction reactions in a polarized
$ep$ collider mode of HERA with respect to their sensitivity to the
proton's polarized gluon distribution. A special emphasis is put on
the `resolved' part of the cross sections which in principle opens the
possibility to determine for the first time also the completely
unknown parton content of longitudinally polarized photons. In the
very promising case of dijet production we also investigate the impact
of parton showering, hadronization and jet finding on the parton level
results.
\end{abstract}

\section{Introduction}
\noindent
Despite the recent experimental progress in the field of
polarized deep-inelastic scattering (DIS), the presently 
available data sets of inclusive and semi-inclusive DIS are still
not sufficient to accurately fix the spin-dependent parton distributions
$\Delta f(x,Q^2)$ of the nucleons, in particular they hardly constrain the 
gluon density $\Delta g$ [1-3]. Even less satisfactory is 
the situation for the $\Delta f^{\gamma}(x,Q^2)$, the parton
distributions of longitudinally (more precisely, circularly) polarized
photons, where nothing at all is known experimentally.

In past years, HERA has already been very successful in pinning down
the proton's unpolarized gluon distribution $g(x,Q^2)$ from
observations of scaling violations in increasingly accurate
$F_2(x,Q^2)$ measurements and from several exclusive processes such as
jet, large-$p_T$ hadron, and heavy flavour production. Moreover, since
the bulk of the HERA events originates from the photoproduction region
($Q^2\rightarrow 0$), `resolved' processes have been measured, in
which the (quasi-real) photon interacts not only in a `direct'
(`point-like') way but resolves into its hadronic structure.  These
HERA measurements [4-13] have been precise enough to improve not only our
knowledge about the parton content of the unpolarized proton, but even
about that of the photon, $f^{\gamma}$.

Given the success of such unpolarized photoproduction experiments at
HERA, it seems to be quite natural to closely examine the same
processes also for the situation with longitudinally polarized beams, in
order to determine their sensitivity to $\Delta g$ {\em{and}} $\Delta
f^{\gamma}$. In preceding studies \cite{wir} we have already analyzed
jet and heavy flavour production, with the latter turning out to be
less useful at collider energies. Here we reanalyze in detail the
whole range of conceivable photoproduction processes, covering this
time also single-inclusive charged hadron, prompt photon, and
Drell-Yan dimuon production. Our previous results for single-inclusive
jet production are extended towards the largest values of the
laboratory rapidity $\eta_{LAB}$ which should be experimentally
accessible. At large $\eta_{LAB}$ one finds an improved sensitivity to
$\Delta f^{\gamma}$.  In the case of dijet production we study an
additional, less exclusive, kinematic selection (as compared to our
previous analysis). This selection allows jets to be measured at higher
transverse energies without loss of statistics, and is also more suited to
eventual next-to-leading order (NLO) QCD calculations.

It should be stressed that HERA could play a unique role in the
determination of the parton content of polarized photons even if it 
should succeed in establishing only the very existence of a resolved
contribution to polarized photon-proton reactions. 
Other determinations of the $\Delta f^{\gamma}$, for instance, via
a measurement of the photon's spin-dependent structure 
function $g_1^{\gamma}$ in polarized $e^+e^-$ collisions, are not planned in
the near future and in case of future lepton-nucleon fixed-target 
photoproduction experiments such as COMPASS \cite{compass},
the resolved component is too small to be extracted experimentally \cite{wir}.

Our previous results, as well as most of those presented here, are
based only on leading order (LO) QCD parton level calculations. Of
course, in real jet production processes, initial and final state QCD
radiation, as well as non-perturbative effects such as hadronization
are also present. These lead to multiparticle final states, and
necessitate the use of a jet algorithm to recombine the final-state
particles into jets. The details of the final-state and the jet
algorithm used can in practice lead to significant differences with
respect to the parton level results.  
In general at HERA for jets of the energies considered here, these
effects are larger than the smearing introduced by detector
resolutions (see, for example \cite{jet2ph}).
In order to investigate the impact of such effects on
the measurement of asymmetries at a polarized HERA, we use a Monte
Carlo simulation, SPHINX \cite{SPHINX}, which includes the polarized
matrix elements for jet photoproduction. Here we limit ourselves to
the case of dijet production which is the only process that provides a
clean separation of the `resolved' part of the cross section, as will
be discussed later. Nevertheless, we expect similar hadronization
effects also for single-inclusive jet or hadron production.
 
Our contribution is organized as follows: In the next section we
collect the necessary ingredients for our calculations. Section 3 is
devoted to a discussion of the various conceivable photoproduction
processes, namely inclusive jet, inclusive hadron, heavy flavour,
prompt photon, and dimuon production. Section 4 contains a more
detailed investigation of dijet production including our Monte Carlo
studies. Finally, section 5 presents the conclusions.

\section{Status of the Polarized Parton Distributions \\ of the
Proton and the Photon}
\noindent
\begin{figure}[th]
\begin{center}
\vspace*{-1.3cm}
\epsfig{file=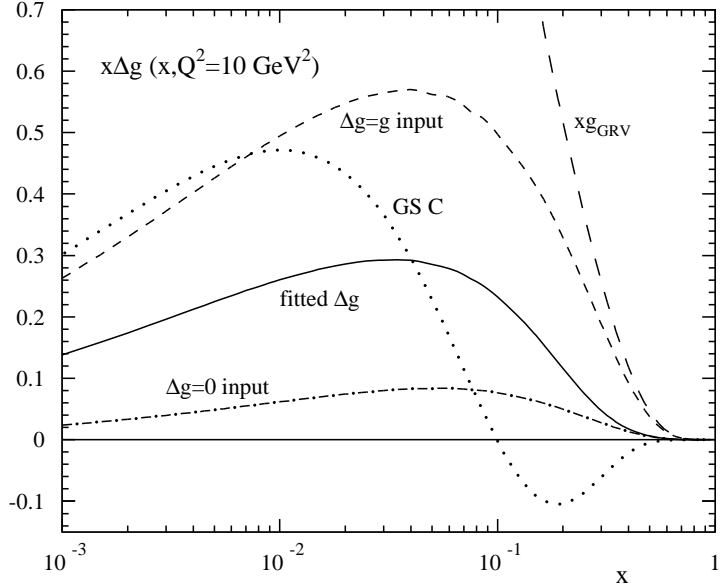,width=11cm}
\vspace*{-1cm}
\caption{\sf Gluon distributions at $Q^2=10\,{\mathrm{GeV}}^2$
of the four LO sets of polarized parton distributions used. 
The dotted line refers to set C of \cite{gs}, whereas the other 
distributions are taken from \cite{grsv} as described in the text. 
Also shown (long-dashed line) is the unpolarized LO gluon distribution 
of \cite{grv}.}
\vspace*{-0.8cm}
\end{center}
\end{figure}
Even though all recent analyses of polarized DIS [1-3,17] have been 
carried out at next-to-leading order (NLO) accuracy, 
we have to stick to LO calculations 
throughout this work since the NLO corrections to polarized photoproduction 
processes are not yet known (except for prompt photon production \cite{gv} 
and the `direct' part of single-inclusive hadron production \cite{dv}).
This implies use of LO parton distributions, which have been provided 
in the studies of \cite{grsv,gs} where various sets, mainly differing in 
the $x$-shape of the polarized gluon distribution, were presented. 

To study the physics potential of HERA for pinning down $\Delta g$ in 
photoproduction processes we will choose four very different LO sets which
represent to a good extent the presently large uncertainty in $\Delta g$.
One should keep in mind that all these sets provide very good descriptions 
of the present polarized DIS data. For definiteness we will take the LO
`standard' set of the radiative parton model analysis \cite{grsv}, which
corresponds to the best-fit result of that paper\footnote{It should be 
noted that one obtains almost indistinguishable results for all 
photoproduction cross sections and asymmetries presented in this
contribution if one uses the best-fit in the so-called `valence'
scenario of \cite{grsv} instead. These two sets mainly differ in
the strange sea-quark content which is of minor importance for
our calculations.}, along with two other `extreme' sets of \cite{grsv} 
which are based on either assuming $\Delta g (x,\mu^2) = g(x,\mu^2)$ or 
$\Delta g(x,\mu^2)=0$ at the low input scale $\mu$ of \cite{grsv}, 
where $g(x,\mu^2)$ is the unpolarized LO GRV \cite{grv} input gluon 
distribution. In the following these two sets will be referred to as 
`$\Delta g=g$ input' and `$\Delta g=0$ input' scenarios, respectively. 
The gluon of set C of \cite{gs}, henceforth denoted as
`GS C', is qualitatively different as it has a substantial negative 
polarization at large $x$. We will therefore also use this set in our 
calculations. For illustration, we show in fig.\ 1 the gluon distributions
of the four different sets of parton distributions we will use, taking a
typical scale $Q^2=10\,{\mathrm{GeV}}^2$. We have also plotted the unpolarized
GRV LO gluon distribution in fig.\ 1, in order to demonstrate that the spin
asymmetry for any process sensitive to the gluon is likely to 
be very small {\em if} it gets sizeable contributions from the region of 
small $x_{gluon}$. This observation has implications for the kinematical
cuts we will choose in our phenomenological studies to be presented in the 
next sections.
\begin{figure}[th]
\begin{center}
\vspace*{-1.4cm}
\epsfig{file=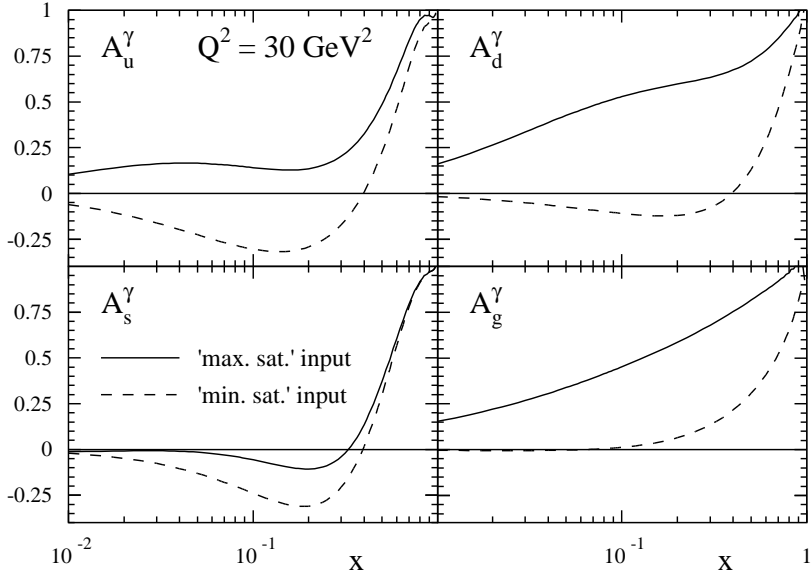,width=12cm}
\vspace*{-0.7cm}
\caption{\sf Photonic LO parton asymmetries 
$A_f^{\gamma}\equiv \Delta f^{\gamma}/f^{\gamma}$ at $Q^2=30\,{\mathrm{GeV}}^2$
for the two scenarios considered in \cite{gvg} (see text). The
unpolarized LO photonic parton distributions were taken from \cite{grvg}.}
\vspace*{-0.8cm}
\end{center}
\end{figure}

In the case of photoproduction the electron just
serves as a source of quasi-real photons which are radiated according
to the Weizs\"{a}cker-Williams spectrum \cite{ww}. The photons can then
interact either directly or via their partonic structure (`resolved'
contribution). For a longitudinally polarized electron beam, the
resulting photon will be also longitudinally (circularly) polarized and, 
in the resolved case, the {\em polarized} parton distributions 
$\Delta f^{\gamma}(x,Q^2)$ of the photon enter the calculations. 
The $\Delta f^{\gamma} (x_{\gamma},Q^2)$ are completely unmeasured so far, 
so that models for them have to be invoked. To obtain a realistic estimate 
for the theoretical uncertainties in the polarized photonic parton densities
two very different scenarios were considered in \cite{gvg} assuming
`maximal' ($\Delta f^{\gamma}(x,\mu^2)=f^{\gamma}(x,\mu^2)$) or `minimal'
($\Delta f^{\gamma}(x,\mu^2)=0$) saturation of the fundamental positivity
constraints (similar to the case of nucleons)
\begin{equation}
\label{eq:pos}
|\Delta f^{\gamma}(x,\mu^2)| \leq f^{\gamma}(x,\mu^2)
\end{equation}
at the input scale $\mu$ for the QCD evolution. 
Here $\mu$ and the unpolarized photon structure functions 
$f^{\gamma}(x,\mu^2)$ were adopted from the phenomenologically successful 
radiative parton model predictions in \cite{grvg}. The results of these 
two extreme approaches are presented in fig.\ 2 in terms of the 
photonic parton asymmetries 
$A_f^{\gamma} \equiv\Delta f^{\gamma}/f^{\gamma}$, 
evolved to $Q^2=30\,{\mathrm{GeV}}^2$ in LO. 
An ideal aim of measurements in a polarized collider mode of HERA 
would of course be to determine the $\Delta f^{\gamma}$ and to see 
which ansatz is more realistic. The sets presented in fig.\ 2, which we
will use in what follows, should in any case be sufficient to study
the sensitivity of the various cross sections to the $\Delta f^{\gamma}$,
but also to see how far they influence a determination of $\Delta g$.

\section{Photoproduction Reactions at Polarized HERA}
\subsection{General Framework}
\noindent
Let us now turn to some technical preliminaries required for our
analyses. First of all, it is useful to define the 
effective polarized parton densities at a scale $M$
in the longitudinally polarized electron by\footnote{We
include here the additional definition
$\Delta f^{\gamma} (x_{\gamma},M^2) \equiv \delta (1-x_{\gamma})$ for the
direct (`unresolved') case.}
\begin{equation}  
\label{eq:elec}
\Delta f^e (x_e,M^2) = \int_{x_e}^1 \frac{dy}{y} \Delta P_{\gamma/e} (y)
\Delta f^{\gamma} (x_{\gamma}=\frac{x_e}{y},M^2) \;
\end{equation}
($f=q,g$), where $\Delta P_{\gamma /e}$ is the polarized 
Weizs\"{a}cker-Williams spectrum for which we will use
\begin{equation}  
\label{eq:weiz}
\Delta P_{\gamma/e} (y) = \frac{\alpha_{em}}{2\pi} \left[
\frac{1-(1-y)^2}{y} \right] \ln \frac{Q^2_{max} (1-y)}{m_e^2 y^2} \; ,
\end{equation}
with the electron mass $m_e$. For the time being, it seems most
sensible to follow as closely as possible the analyses successfully
performed in the unpolarized case, which implies to introduce the same
kinematical cuts. As in \cite{jet1ph,jet2ph,dij94,chph,ihzeus} we will use
an upper cut\footnote{In H1 analyses of HERA photoproduction data
\cite{jet1h1,jet2h1,chh1,ihh1} the cut $Q^2_{max}=0.01\,{\mathrm{GeV}}^2$
is used along with slightly different $y$-cuts as compared to the
corresponding ZEUS measurements \cite{jet1ph,jet2ph,dij94,chph,ihzeus}, 
which leads to smaller rates.} $Q^2_{max}=4\,{\mathrm{GeV}}^2$, 
and the $y$-cuts $0.2 \leq y \leq 0.85$ 
(for single-jet \cite{jet1ph} and heavy flavour \cite{chph} production) 
and $0.2 \leq y \leq 0.8$ (for dijet \cite{jet2ph,dij94}, 
inclusive hadron \cite{ihzeus}, prompt
photon \cite{dirgam} and Drell-Yan dimuon production) will be imposed.

Furthermore, it should be noted that in what follows a polarized cross 
section will always be defined as
\begin{equation}
\Delta \sigma \equiv \frac{1}{2} \left( \sigma (++)-\sigma (+-) \right) \; ,
\end{equation}
the signs denoting the helicities of the scattering particles. The
corresponding unpolarized cross section $\sigma$ is obtained by taking the sum
instead, and the experimentally relevant 
cross section asymmetry is $A\equiv \Delta \sigma/\sigma$.
Whenever calculating an asymmetry $A$, we will
use the LO GRV parton distributions for the proton \cite{grv} and the
photon \cite{grvg} to calculate the unpolarized cross section.
For consistency, we will employ the LO expression for the strong coupling
$\alpha_s$ with \cite{grsv,gs,grv,gvg,grvg} $\Lambda_{QCD}^{(f=4)}=200$ MeV 
for four active flavours.

Rapidity distributions will always be presented in terms of the laboratory
frame rapidity $\eta_{LAB}$ which is related to the centre-of-mass 
$\eta_{cms}$ via 
\begin{equation}
\eta_{cms} = \eta_{LAB} - \frac{1}{2} \ln (E_p/E_e)\;\;. 
\end{equation}
As conventional for HERA, $\eta_{LAB}$ is defined to be positive in the 
proton forward direction. Studying distributions in $\eta_{LAB}$ is a 
particularly suitable way of separating the direct from the resolved 
contributions in single-inclusive measurements of jets or
hadrons: for negative $\eta_{LAB}$, the main contributions are expected 
to come from the region $x_{\gamma} \rightarrow 1$ and thus mostly from the 
direct piece at $x_{\gamma}=1$. The situation is reversed at positive 
$\eta_{LAB}$.

In the asymmetry plots we will always show the expected statistical errors 
$\delta A$ for a measurement at HERA, estimated from
\begin{equation}  
\label{eq:errors}
\delta A = \frac{1}{P_e P_p \sqrt{{\cal L} \sigma \epsilon}} \; ,
\end{equation}
where $P_e$, $P_p$ are the beam polarizations, ${\cal L}$ is the integrated
luminosity and $\epsilon$ the detection efficiency for the 
desired final state. For all our calculations we assume $P_e * P_p=0.5$ and
a conservative value of ${\cal L}=100$/pb. 
%
\subsection{Single-Inclusive Jet Production}
%
For this process we can be rather brief, as all details of the calculation
were already described in our previous report \cite{wir}. Here we only
present updated plots of the cross sections and their asymmetries, 
extending to larger, but still experimentally accessible, values of 
$\eta_{LAB}$.
\begin{figure}[th]
\begin{center}
\vspace*{-0.9cm}
\epsfig{file=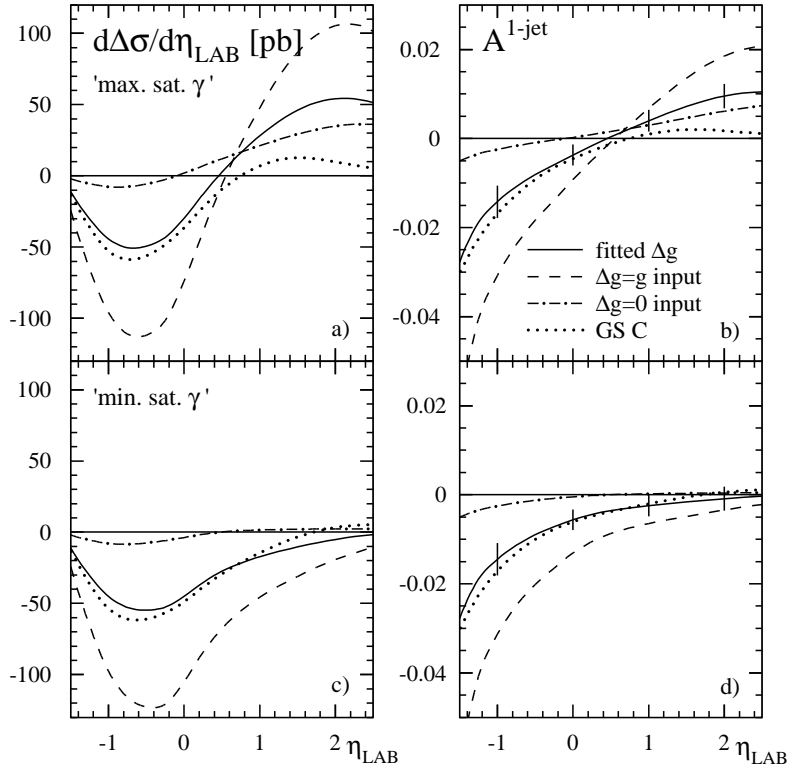,width=11.5cm}
\vspace*{-0.8cm}
\caption{\sf {\bf a:} $\eta_{LAB}$ dependence of the polarized single-jet
inclusive photoproduction cross section in $ep$-collisions at HERA, integrated
over $p_T > 8$ GeV. The renormalization/factorization scale was
chosen to be $M=p_T$. The resolved contribution to the cross section has
been calculated with the `maximally' saturated set of polarized photonic
parton distributions. {\bf b:} Asymmetry corresponding to {\bf a}. The
expected statistical errors have been calculated according to \
(\ref{eq:errors}) and as described in the text. {\bf c,d:} Same as {\bf a,b},
but for the `minimally' saturated set of polarized photonic parton
distributions.}
\vspace*{-0.7cm}
\end{center}
\end{figure}

Fig.\ 3 shows our results for the single-inclusive jet cross section 
and its asymmetry vs.\ $\eta_{LAB}$ and integrated over $p_T>8$ GeV 
for the four sets of the polarized proton's parton distributions. 
For figs.\ 3a,b we have used the `maximally'
saturated set of polarized photonic parton densities, whereas figs.\ 3c,d
correspond to the `minimally' saturated one. Comparison of figs.\ 3a,c or
3b,d shows that as expected the direct contribution clearly dominates for
$\eta_{LAB} \lesssim -0.5$, where also differences between the polarized gluon
distributions of the proton show up clearly. Furthermore, the cross sections
are generally large in this region with asymmetries of a few per cent. At
positive $\eta_{LAB}$, in particular in the region $\eta_{LAB}\gtrsim 2$ now
included in the plot, we find that the cross section is dominated by the 
resolved contribution. Here it is sensitive to the parton content 
of both the polarized proton {\em and} the photon, which means that one can 
only learn something about the polarized photon structure functions if 
the polarized parton distributions of the proton are
already known to some accuracy, e.g., from an analysis of the region
of negative rapidities. We note that the dominant
contributions to the resolved part at large $\eta_{LAB}$ are driven by the
polarized photonic {\em gluon} distribution $\Delta g^{\gamma}$.

We have included in the asymmetry plots in figs.\ 3b,d the expected
statistical errors $\delta A$ at HERA estimated from eq.\ (\ref{eq:errors})
for $\epsilon=1$. From the results it appears that a measurement of 
the proton's $\Delta g$ should be possible from single-jet events at 
negative rapidities where the contamination from the resolved contribution 
is minimal.
\begin{figure}[th]
\begin{center}
\vspace*{-0.9cm}
\epsfig{file=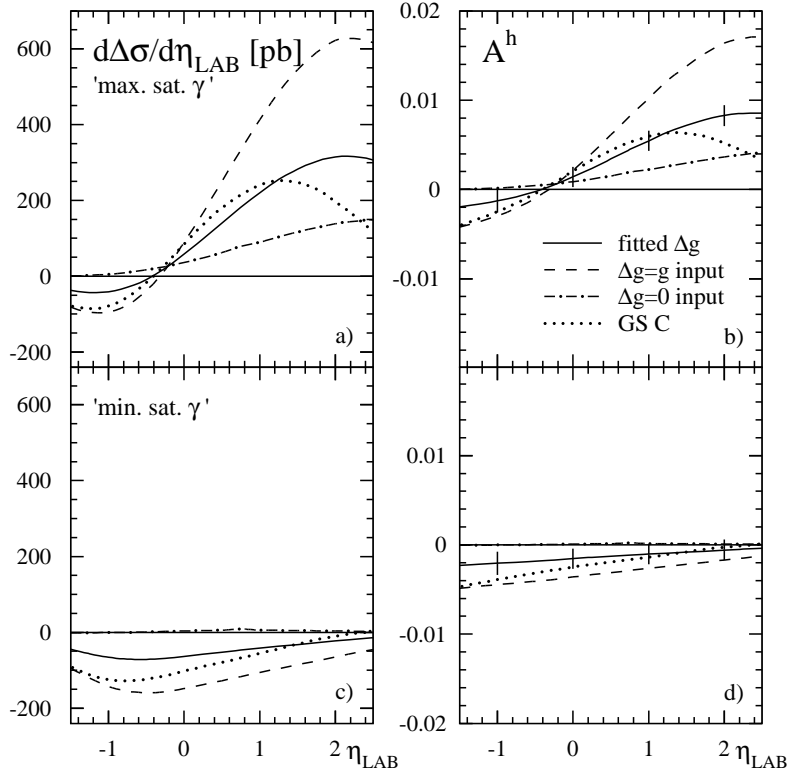,width=11.5cm}
\vspace*{-0.9cm}
\caption{\sf {\bf a:} $\eta_{LAB}$ dependence of the polarized 
single-inclusive hadron photoproduction cross section in 
$ep$-collisions at HERA, integrated over $p_T > 2$ GeV. The error bars 
have been calculated assuming an efficiency $\epsilon=0.8$ for 
the hadron detection. All other parameters were chosen as in fig.\ 3.} 
\vspace*{-0.8cm}
\end{center}
\end{figure}

%
\subsection{Single-Inclusive Charged Hadron Production}
%
\hspace*{0.0001mm}From our results for one-jet production in 
fig.\ 3 it seems worthwhile to consider also single-inclusive 
production of charged hadrons. 
At first glance, this process appears less interesting than jet production, 
as the cross section for producing a definite hadron at a given $p_T$ will 
always be smaller than the one for a jet. On the other hand, in case 
of inclusive hadrons one can obviously go experimentally to $p_T$ much 
smaller than the $p_T^{min}=8$ GeV employed in our single jet study. 
Moreover, in the unpolarized case single-inclusive hadron production 
was successfully studied experimentally at HERA prior to 
jets \cite{ihh1,ihzeus}. 

The generic LO cross section formula for the photoproduction of a single 
hadron $H$ with transverse momentum $p_T$ and cms-rapidity $\eta$ 
in polarized $ep$ collisions reads:
\begin{equation} 
\label{eq:wqc}
\frac{d^2 \Delta \sigma^H}{dp_T d\eta} =  \sum_{f^e,f^p,c}
\Delta f^e (x_e,M^2) \otimes \Delta f^p (x_p,M^2) \otimes
\frac{d^2 \Delta \hat{\sigma}^{f_e f_p \rightarrow cd}}{dp_T d\eta} 
\otimes D_c^H (z,M^2) \; ,
\end{equation}
where $\otimes$ denotes a convolution and the sum is running over all
appropriate $2\rightarrow 2$ subprocesses for the direct
($\gamma b\rightarrow cd$, $\Delta f^e (x_e,M^2) \equiv
\Delta P_{\gamma/e}(x_e)$) and resolved ($ab\rightarrow cd$) cases. 
These subprocesses are the same as the ones for jet production.
In (\ref{eq:wqc}), $\hat{s} \equiv x_e x_p s$ and 
$M$ is the factorization/renormalization scale for which we will 
use $M=p_T$ as for our jet calculations. 
The $\Delta f^p$ stand for the polarized parton distributions 
of the proton, and the $D_c^H$ are the unpolarized fragmentation 
functions for $c\rightarrow H$. For the latter we will use the LO 
functions of \cite{bkk} which yield a good description of the 
unpolarized HERA inclusive hadron data \cite{ihh1,ihzeus}. 
Needless to say that we obtain the corresponding unpolarized 
LO jet cross section $d^2 \sigma^H/dp_T d\eta$ by using LO unpolarized 
parton distributions and subprocess cross sections in (\ref{eq:wqc}).

Figs.\ 4a,b show our results for the sum of charged pions and kaons 
after integration over $p_T>2$ GeV, where all other parameters were 
chosen exactly as for figs.\ 3a,b. One can see that the spin asymmetries 
behave rather similarly in shape as the corresponding results
in figs.\ 3a,b, but are somewhat smaller in magnitude mainly due to the
smaller $x$ values probed here for $p_T^{min}=2$ GeV compared to
the single jet case where $p_T^{min}=8$ GeV (see also fig.\ 1).
Nevertheless, the expected statistical errors, 
calculated for the realistic choice
of $\epsilon=0.8$ in eq.\ (\ref{eq:errors}) and displayed in fig.\ 4b,d, 
demonstrate that single-inclusive hadron photoproduction is also a very 
promising candidate.

We finally mention that apart from charged-hadron production we have also 
studied the photoproduction of $\Lambda$ baryons. This is another particularly
interesting topic, as it turns out to be possible experimentally to 
determine the polarization of the produced $\Lambda$'s from their 
parity-violating decays to $p\pi$. This implies that a measurement of the 
spin asymmetry in this case could allow statements concerning the 
{\em spin-dependent} fragmentation functions of the $\Lambda$, hereby
possibly providing some complementary new insight into `spin-physics'. 
Encouraging theoretical predictions for photoproduction
of $\Lambda$'s are also presented in these proceedings \cite{dsv}. 

\subsection{Other Processes: Heavy Flavours, Prompt Photons, and Drell-Yan
dimuon production}
%
Heavy flavour (charm) production was already considered in our previous 
report \cite{wir}, where we found expected statistical 
errors roughly of the size of
the spin asymmetry itself. This process, albeit being very sensitive to
$\Delta g$ thanks to the dominance of the photon-gluon-fusion subprocess
$\gamma g\rightarrow c\bar{c}$, appears therefore less favourable than 
the ones considered in the previous subsections. It is expected to be 
more useful for fixed-target experiments, where measurements of 
$\Delta g$ in open-charm production are planned \cite{compass}.

Prompt photon production at HERA is currently under investigation 
in the unpolarized case \cite{dirgam}. Here important contributions at LO
result, for instance, from the subprocess $qg \rightarrow \gamma q$,
with the initial quark coming from the resolved photon and the gluon
from the proton, or vice versa. This process, which is therefore 
sensitive to $\Delta g$ in the polarized case, has to compete with 
$q\bar{q}\rightarrow \gamma g$ annihilation and with the Compton process 
$\gamma q\rightarrow \gamma q$ from the direct (unresolved) photon. 
We found that the main problem of prompt photon production is the
smallness of the unpolarized cross section (with respect to, say, the
single-inclusive hadron cross section), which results in large expected 
statistical errors, $\delta A^{\gamma}> A^{\gamma}$. The experimental 
study of prompt photon production at a polarized HERA thus appears
to be impossible.

Finally, we have also examined photoproduction of Drell-Yan dimuon pairs.
This process is of potential interest due to the fact that
the relevant LO subprocess is simply $q\bar{q} \rightarrow \mu^+ \mu^-$,
which means for the case of photoproduction that only the resolved 
contribution is present at LO. The experimental finding of a nonvanishing 
asymmetry would therefore be evidence for the existence of polarized
(anti)quarks in a polarized real photon. Unfortunately, again the 
expected statistical errors turn out to be way larger than the 
asymmetry itself due to the smallness of the unpolarized cross section.

\section{Dijet Production}
\def\rsep{R_{\rm sep}} 
\def\cts{\cos\theta^{\ast}}
\def\xgo{x_\gamma^{OBS}} 
\def\xpo{x_p^{OBS}} \def\xg{x_\gamma}
\def\ETJ{E_T^{jet}} 
\def\ETCJ{E_T^{CALJet}} 
\def\ETTJ{E_T^{TLTJet}}
\def\ETJM{E_T^{min}} 
\def\ETAJ{\eta^{jet}} 
\def\PHIJ{\phi^{jet}}
\def\ETACJ{\eta^{CALJet}} 
\def\ETATJ{\eta^{TLTJet}}
\def\ETAB{\bar{\eta}} 
\def\EEP{E^\prime_{e}}
\def\TEP{\theta^\prime_{e}} 
\def\MJJ{M_{jj}} 
\def\DETA{\Delta\eta}
\def\ptmin{\hat{p}_T^{\rm min}} 

In the case of unpolarized photoproduction of dijets at HERA, an
experimental criterion for a distinction between direct and resolved
contributions has been introduced \cite{jeff,jet2ph,dij94}. In our previous
analysis \cite{wir}, we have adopted this criterion for the polarized
case to see whether it would enable a further access to $\Delta g$
and/or the polarized photon structure functions. Very encouraging
results were obtained.  However, the results presented in \cite{wir}
were based on LO QCD parton level calculations. 
One implication of this is that, although
the kinematic cuts applied to jets in \cite{wir} reflect the cuts
which have been used in experimental analyses, the `jets' to which
they are applied consist of single quarks or gluons. In a real event,
higher order effects such as QCD radiation (in the initial and final
state) and non-perturbative effects such as hadronization are also
present. These lead to multiparticle final states, and necessitate the
use of a jet algorithm to recombine the final state particles into
jets. The jet kinematics should then reflect the kinematics of the
initiating parton. However, the details of the final state and the jet
algorithm used can in practice lead to significant differences with
respect to the parton level predictions. In general, at HERA these
effects are larger than smearing introduced by 
detector resolutions (see, e.g., \cite{jet2ph}).

In this section, we will therefore present a more sophisticated
analysis of dijet production that includes these additional effects,
as modelled in the Monte Carlo simulation mentioned in the
introduction. Among all our results, the ones presented in this
section are therefore the `most realistic' ones. We expect the typical
size of the effects from hadronization and QCD radiation 
found here to be representative for all other photoproduction 
processes considered in this work.

\subsection{Separating Direct and Resolved Contributions}
%
To begin with, let us briefly recall the underlying idea for the 
separation of direct and resolved contributions \cite{jeff,jet2ph,dij94}. 
The generic expression for 
the polarized cross section for the photoproduction of two jets with 
laboratory system rapidities $\eta_1$, $\eta_2$ is to LO
\begin{equation} 
\label{eq:wq2jet}
\frac{d^3 \Delta \sigma}{dp_T d\eta_1 d\eta_2} = 2 p_T
\sum_{f^e,f^p} x_e \Delta f^e (x_e,M^2) x_p \Delta f^p (x_p,M^2)
\frac{d\Delta \hat{\sigma}}{d\hat{t}} \; ,
\end{equation}
where $p_T$ is the transverse momentum of one of the two jets (which balance
each other in LO) and
\begin{equation}
x_e \equiv \frac{p_T}{2 E_e} \left( e^{-\eta_1} + e^{-\eta_2} \right)\;\; , \;
x_p \equiv \frac{p_T}{2 E_p} \left( e^{\eta_1} + e^{\eta_2} \right) \; .
\end{equation}
The important point is that measurement of the jet rapidities allows
for fully reconstructing the kinematics of the underlying hard subprocess
and thus for determining the variable \cite{jet2ph}
\begin{equation}
x_{\gamma}^{OBS} = \frac{\sum_{jets} p_T^{jet} e^{-\eta^{jet}}}{2yE_e} \; ,
\end{equation}
which in LO equals $x_{\gamma}=x_e/y$, with $y$ as before being the
fraction of the electron's energy taken by the photon. Thus it becomes
possible to experimentally select events at large $x_{\gamma}$,
$x_{\gamma}^{OBS} > 0.75$ \cite{jeff,jet2ph,dij94}, 
hereby extracting the {\em direct} 
contribution to the cross section with a relatively small contamination 
from resolved processes. Conversely, the events 
with $x_{\gamma}^{OBS}\leq 0.75$ 
will represent the resolved part of the cross section. This procedure 
should therefore be ideal to extract $\Delta g$ on the one hand, and 
to examine the polarized photon structure functions on the other.

\subsection{Monte Carlo Simulation}
%
In order to investigate the impact of effects from hadronization on the 
measurement of spin asymmetries at a polarized HERA, a Monte Carlo simulation,
SPHINX \cite{SPHINX}, has been used. SPHINX is based upon the general
purpose library of Monte Carlo routines, PYTHIA \cite{PYTHIA}.  

As used in this analysis, the program generates hard scatterings
according to the LO QCD matrix element, in addition it includes
initial and final state QCD radiation calculated to leading
logarithmic accuracy in $Q^2$. The parton shower evolves from the
scale of the hard interaction down to $1\,{\mathrm{GeV}}^2$. Below
this scale, the further evolution is considered to be
non-perturbative, and the Lund string model is used to hadronize the
(in general many) partons.

The additional feature of SPHINX as compared to PYTHIA is that it
includes the polarized matrix elements for jet photoproduction, as
well as implementing the various different polarized parton
distributions (for the proton and photon) discussed previously. Thus
the simulation of asymmetries in realistic final states is
possible. However, the polarization information is only used in the
hard subprocess. The parton showers and fragmentation stages do not
preserve a memory of the polarization.

SPHINX/PYTHIA use $p_T$ for the hard scale of the interaction, which is
the relative transverse momentum of the scattered
partons. Events were generated down to a minimum $p_T$ of 
$5\,{\mathrm{GeV}}$.

As a case study, $50\,{\mathrm{pb}}^{-1}$ 
of luminosity was generated for direct
and resolved photoproduction for four options, using the GS C parton
distribution for the proton and the maximally and minimally saturated
distributions for the photon (36 million events in total). This leaves
a pessimistic (or conservative) estimate of the statistical errors
which might eventually be obtained, but is sufficient to show whether
or not the migrations between parton and hadron level might preserve
observable asymmetries.

Most measurements of jet cross sections at hadron-hadron colliders and
in photoproduction at HERA have used some variation of a cone-based
jet algorithm. These algorithms maximize the transverse energy inside
a cone in $\eta-\phi$ space. They are invariant under boosts along the
beam axis, and successfully isolate `hard' physics which in
hadron-hadron or photon-hadron interactions is generally at large
transverse energies with respect to the beam axis. However, they
suffer from various ambiguities in how jet finding is initiated in the
first place and in how jets are merged or split, and are in general
not infrared safe \cite{safe}. 

In this analysis the cluster algorithm KTCLUS \cite{Cat93,Ell93} is
therefore used. In this algorithm, the quantity
\begin{equation}
d_{i,j}=\left((\eta_i-\eta_j)^2+(\phi_i-\phi_j)^2\right) 
{\rm min}(E_{T_i},E_{T_j})^2
\end{equation}
is calculated for each pair of objects (where the initial objects are
the final state particles), and for each individual object:
\begin{equation}
d_i=E^2_{T_i}.
\end{equation}
If, of all the numbers $\{ d_{i,j},d_i \}$, $d_{k,l}$ is the smallest, then
objects $k$ and $l$ are combined into a single new object.  If however
$d_k$ is the smallest, then object $k$ is a jet and is removed from
the sample.  This is repeated until all objects are assigned to jets.
The parameters for the jet are calculated
as:
\begin{equation}
\label{jetpar_eqn}
\ETJ = \sum_i E_{T_i}\;\;,\;\;
\eta^{jet} = \frac{1}{\ETJ}\sum_i E_{T_i} \eta_i\;\;,\;\;
\phi^{jet} = \frac{1}{\ETJ{}}\sum_i E_{T_i} \phi_i
\end{equation}
in which the sums run over all particles belonging to the jet.  This
scheme is also used to determine the parameters of the intermediate
objects.  In this scheme KTCLUS retains the advantages of cone
algorithms, without suffering from their associated problems. This
algorithm has now been used in more recent experimental analyses 
(see, e.g., \cite{dij94}).
\begin{figure}[th]
\begin{center}
\vspace*{-1.4cm}
\epsfig{file=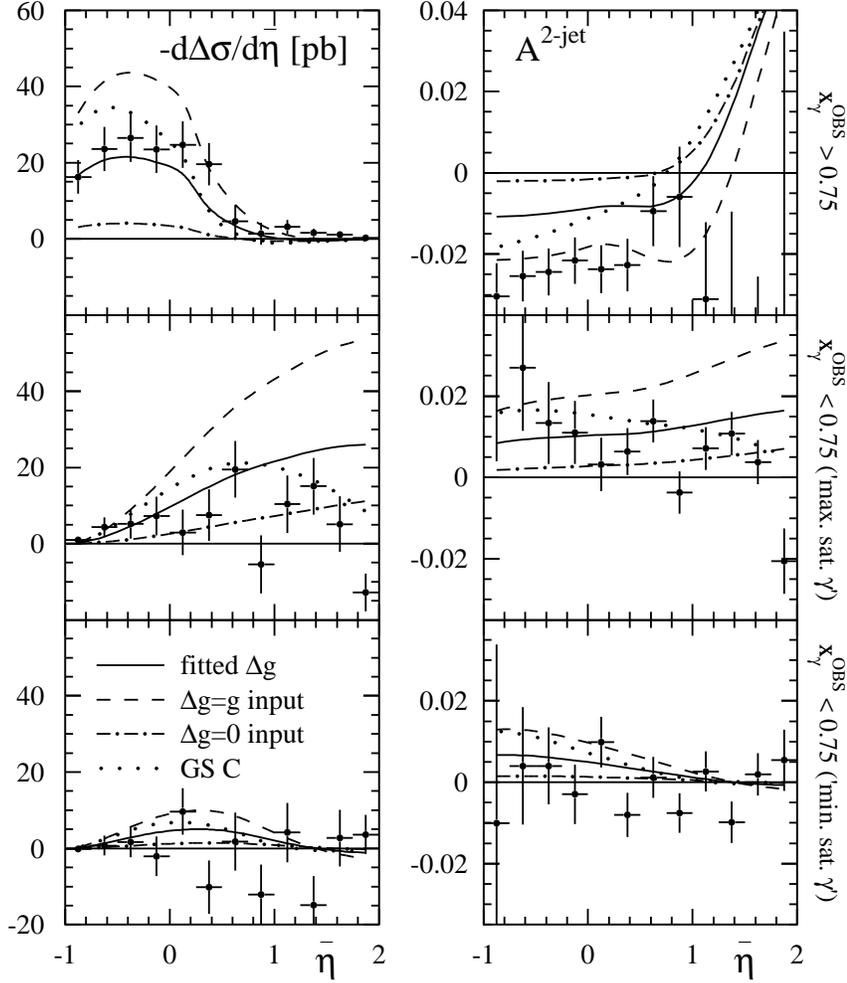,width=12.5cm}
\vspace*{-1.2cm}
\caption{\sf $\bar{\eta}$-dependence of the `direct' and `resolved' parts
of polarized two-jet photoproduction at HERA for the four
different sets of polarized parton densities of the proton. The left hand 
column shows the cross section for $x_{\gamma}^{OBS}>0.75$, and
for $x_{\gamma}^{OBS}<0.75$ with the `maximally' and `minimally' saturated
sets of polarized photonic parton densities, reading from top to bottom,
respectively. The right hand column shows the corresponding asymmetries.
In the top row,
the resolved contribution with $x_{\gamma}^{OBS}>0.75$ has been included
using the `maximal' photon set. 
Also shown are the Monte Carlo results based on a generated sample of 50/pb 
using the GS C \cite{gs} distributions (see text).} 
\vspace*{-0.8cm}
\end{center}
\end{figure}

\subsection{Results for Dijet Production}
We first compare the Monte Carlo results with the predictions
presented in our previous study. For this purpose we integrate over
the cross section in (\ref{eq:wq2jet}) to obtain $d\Delta
\sigma/d\bar{\eta}$, where $\bar{\eta} \equiv (\eta_1 +
\eta_2)/2$. Furthermore, we will apply the cuts \cite{jet2ph} $|\Delta
\eta| \equiv |\eta_1-\eta_2| \leq 0.5 \; , \;\; p_T>6\,\mbox{GeV}$.
The results are shown in fig.\ 5. The left hand column shows the
$d\Delta\sigma/d\ETAB$ distribution for $\xgo > 0.75$, 
and for $\xgo <0.75$ with the `maximally' and `minimally' saturated 
sets of polarized photonic parton distributions, reading
from top to bottom, respectively. The right hand column shows the
corresponding asymmetries. 
The bin widths are chosen to be commensurate with the experimental
resolution, rather than being optimized for the size of the 
statistical sample shown here. 
\begin{figure}[th]
\begin{center}
\vspace*{-0.9cm}
\epsfig{file=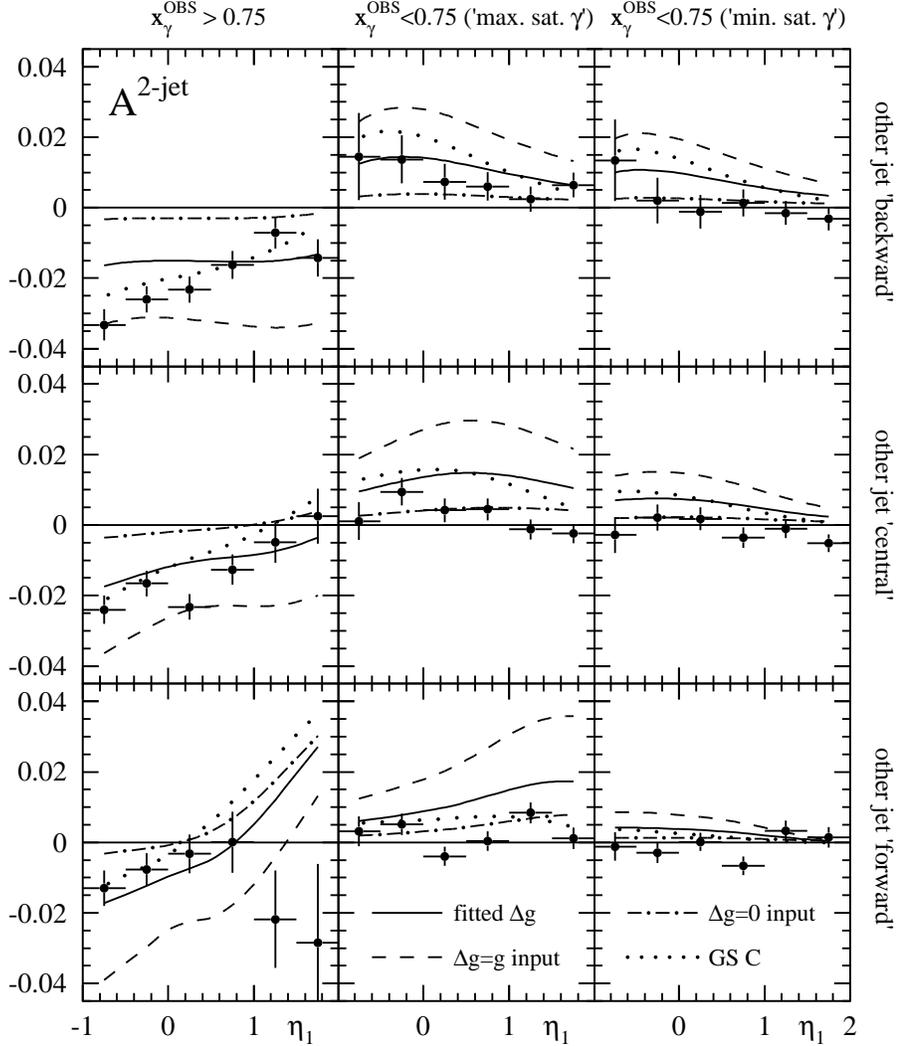,width=13cm}
\vspace*{-1.0cm}
\caption{\sf Similar as in fig.\ 5 but now showing the 
$\eta_1$-dependence of the polarized two-jet asymmetries 
at HERA for three different kinematical configurations (see text). } 
\vspace*{-0.8cm}
\end{center}
\end{figure}

In the high $\xgo$ case ($\xgo>0.75$), the agreement with the 
parton level predictions is generally good in both the cross section 
and the asymmetry. The
exception is at very forward rapidities ($\eta > 1$) where the parton
level cross section is very small and the behaviour of the dijet
asymmetry is dominated by the effects of high $\xgo$ resolved
contributions and the migration of LO direct jets towards high $\eta$
due to the presence of the proton remnant (to which they are
colour-connected).  Nevertheless, overall the prospects for an
independent, direct constraint on the polarized parton distributions
in the proton by this method appear to be good.  In the equally
interesting low $\xgo$ case ($\xgo<0.75$), the agreement is not 
as good, although it is still reasonable. 
In particular, the minimally saturated case
clearly still shows lower asymmetries than the maximally saturated
one, and is in fact consistent with zero for this luminosity.
\begin{figure}[th]
\begin{center}
\vspace*{-0.9cm}
\epsfig{file=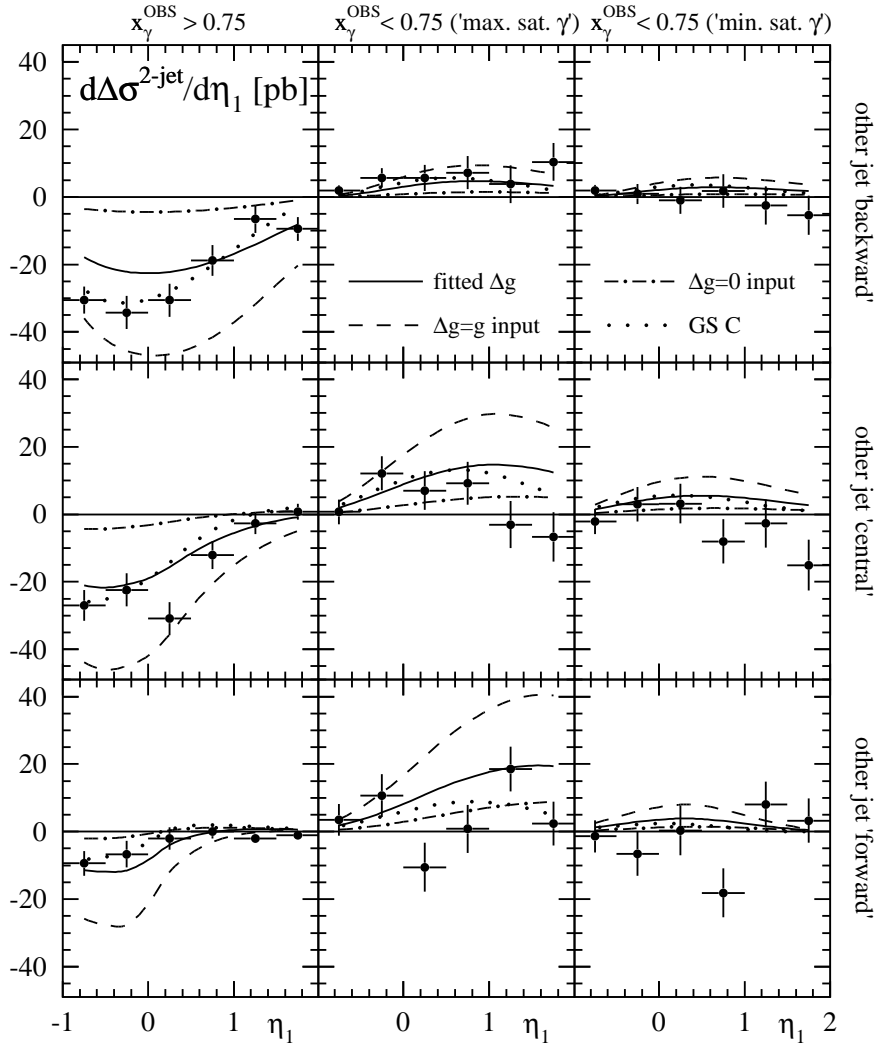,width=13cm}
\vspace*{-1.0cm}
\caption{\sf Same as in fig.\ 6 but now showing the corresponding
polarized two-jet cross sections.} 
\vspace*{-0.8cm}
\end{center}
\end{figure}

In the unpolarized case, the above cross sections were proposed 
following arguments based upon LO QCD \cite{jeff}, but have since been 
calculated to NLO \cite{kk}. In the course of these calculations, it was
discovered that soft gluon radiation leads to difficulties when the
$\ETJ$ threshold is the same for both jets at NLO, and so subsequent
measurements have concentrated on cases where a different threshold is
applied to each jet. We study here an example of such a cross
section. In this case we demand that there must be two jets with $\ETJ
> 6$ GeV, as before, but that at least one of them must have $\ETJ > 8$
GeV. The jets are then symmetrized in $\eta$. The measured cross
sections are then $d\sigma/d\eta_1^{jet}$ for one jet, where the other jet is
required to be either `backward' ($-1 < \eta_2^{jet} < 0$), `central' ($0 <
\eta_2^{jet} < 2$), or `forward' ($1 < \eta_2^{jet} < 2$). The cross 
sections are measured for the two regions, $\xgo > 0.75$ and $\xgo < 0.75$. 
The $\DETA$ cut is removed. In the LO calculations shown here, which are
currently all that is available for such cross sections in the polarized case,
both jets of course have the same $\ETJ$ anyway, and so both have $\ETJ > 8$
GeV. However, in the MC there is smearing due to the additional
effects which can lead to jets having different $\ETJ$.

The results are shown in fig. 6 (asymmetries) and 7 (cross sections).
Again in the high $\xgo$ case the agreement is good.
In the low $\xgo$
case, the asymmetries are in general lower than in the parton level
case, and indeed are consistent with zero for the minimally saturated
photonic parton distributions. However, there is still a measurable 
non-zero asymmetry in the maximally saturated case. The bins widths in 
these results are dictated by the statistics. 

\section{Summary and Conclusions}
\noindent
We have analyzed various conceivable photoproduction experiments in the 
context of a polarized $ep$-collider mode of HERA. Leading order theory 
predictions for jet and single-inclusive hadron production show a very 
encouraging sensitivity to the polarized gluon distribution of the proton 
and also to the completely unknown parton content of a circularly polarized 
photon. In particular, it turns out that for these processes the differences 
between results obtained for different sets of polarized parton 
distributions are usually clearly larger than the expected statistical errors
for a measurement at polarized HERA, even for a rather conservative
luminosity of 100/pb. In contrast to this, the experimental study of 
photoproduction of heavy flavours, prompt photons and Drell-Yan dimuon pairs 
at a polarized HERA appears impossible. 

In the very promising case of dijet photoproduction which allows for
a separation of the `direct' and `resolved' contributions on an
experimental basis, we have investigated the 
impact of parton showering, hadronization and jet finding on the parton level
results, which is a crucial ingredient for making contact between 
theoretical estimates based on LO parton level calculations and 
the `real world'. We have found that although the combined
effects of hadronization and QCD radiation can be significant,
measurable asymmetries are preserved. Thus the prospects both of obtaining an
independent constraint on the polarized parton distributions in the
proton, and of observing for the first time a possible asymmetry in
those of the photon seem excellent, even with relatively limited
integrated luminosity.

Finally, it should be again stressed that HERA could be {\em{the}} place to
obtain at least some experimental information on the presently 
completely unmeasured parton distributions of a longitudinally polarized
photon in the foreseeable future.
Most of the various photoproduction processes studied here appear to be
promising and experimentally feasible to achieve this aim even with a
conservative luminosity of 100/pb.

 

\begin{thebibliography}{999} 
%
\bibitem{grsv} M.\ Gl\"{u}ck, E.\ Reya, M.\ Stratmann, and W.\ Vogelsang, 
{\it Phys. Rev.} {\bf D53} (1996) 4775.
%
\bibitem{gs} T.\ Gehrmann and W.J.\ Stirling, {\it Phys. Rev.} {\bf D53}
(1996) 6100.
%
\bibitem{ref3} D.\ de Florian, talk presented at the workshop `Deep
Inelastic Scattering off Polarized Targets: Theory Meets Experiment',
Zeuthen, Germany, 1997, to appear in the proceedings; M.\ Stratmann, ibid.
%
\bibitem{jet1h1} I.\ Abt et al., H1 collab., {\it Phys. Lett.} {\bf{B314}}
(1993) 436; C.\ Adloff et al., H1 collab., {\tt DESY 97-179},
{\tt hep-ex/9709017}.
%
\bibitem{jet1ph}  M.\ Derrick et al., ZEUS collab., {\it Phys. Lett.} 
{\bf{B342}} (1995) 417. 
%
\bibitem{jet2h1} T.\ Ahmed et al., H1 collab., {\it Nucl. Phys.} {\bf{B445}}
(1995) 195; C.\ Adloff et al., H1 collab., {\tt DESY 97-164},
{\tt hep-ex/9709004}.
%
\bibitem{jet2ph} M.\ Derrick et al., ZEUS collab., {\it Phys. Lett.} 
{\bf{B322}} (1994) 287; {\bf{B348}} (1995) 665.
%
\bibitem{dij94} J. Breitweg et al, ZEUS collab., {\tt DESY 97-196}, 
{\tt hep-ex/9710018}.
%
\bibitem{chph} M.\ Derrick et al., ZEUS collab., {\it Phys. Lett.} 
{\bf{B349}} (1995) 225; {\bf{B401}} (1997) 192.
%
\bibitem{chh1} S.\ Aid et al., H1 collab., {\it Nucl. Phys.} {\bf B472} (1996)
32.
%
\bibitem{ihh1} I.\ Abt et al., H1 collab., {\it Phys. Lett.} {\bf B328} (1994)
176.
%
\bibitem{ihzeus} M.\ Derrick et al., ZEUS collab., {\it Z. Phys.} {\bf C67}
(1995) 227.
%
\bibitem{dirgam} S.\ Aid et al., H1 collab., paper submitted to the 
`International Europhysics Conference on High Energy Physics', 1997,
Jerusalem, Israel; J.\ Breitweg et al., ZEUS collab., {\tt DESY 97-146},
{\tt hep-ex/9708038}.
%
\bibitem{wir} M. Stratmann and W. Vogelsang, {\it Z. Phys.} {\bf C74} (1997)
641; Proc. of the 1995/96 workshop on `Future Physics at HERA',
Hamburg, Germany, eds. G.\ Ingelman, A.\ De Roeck, and R.\ Klanner, p.\ 815.
%
\bibitem{compass} G.\ Baum et al., COMPASS collab., {\tt CERN/SPSLC 96-14}.  
%
\bibitem{SPHINX} S.\ G\"{u}llenstern et al., {\tt hep-ph/9612278};
O.\ Martin, M.\ Maul, and A.\ Sch\"{a}fer,\\ {\tt hep-ph/9710381}; see also,
O.\ Martin, the SPHINX homepage,\\
{\tt http://www.th.physik.uni-frankfurt.de/~martin/sphinx.html}.
%
\bibitem{abfr} R.D.\ Ball, S.\ Forte, and G.\ Ridolfi, {\it Phys. Lett.}
{\bf B378} (1996) 255; G.\ Altarelli, R.D.\ Ball, S.\ Forte, and G.\ Ridolfi,
{\it Nucl. Phys.} {\bf B496} (1997) 337. 

\bibitem{gv} L.E.\ Gordon and W.\ Vogelsang, {\it Phys. Rev.} {\bf D48} 
(1993) 3136; A.P.\ Contogouris, B.\ Kamal, Z.\ Merebashvili, and 
F.V.\ Tkachov, {\it Phys. Lett.} {\bf B304} (1993) 329; 
{\it Phys. Rev.} {\bf D48} (1993) 4092; {\bf D54} (1996) 7081 (E).
%
\bibitem{dv} D.\ de\ Florian and W.\ Vogelsang, {\tt CERN-TH/97-280}, in
preparation.
%
\bibitem{grv} M.\ Gl\"{u}ck, E.\ Reya, and A.\ Vogt, {\it Z. Phys.}
{\bf{C67}} (1995) 433.
%
\bibitem{ww} C.F. von Weizs\"{a}cker, {\it Z. Phys.} {\bf 88}, 612 (1934);
E.J. Williams, {\it Phys. Rev.} {\bf 45}, 729 (1934).
%
\bibitem{gvg} M. Gl\"{u}ck and W. Vogelsang, {\it Z. Phys.} {\bf C55} (1992)
353; {\bf C57} (1993) 309; M. Gl\"{u}ck, M. Stratmann, and W. Vogelsang, 
{\it Phys. Lett.} {\bf B337} (1994) 373.
%
\bibitem{grvg} M.\ Gl\"{u}ck, E.\ Reya, and A.\ Vogt, {\it Phys. Rev.} 
{\bf D46} (1992) 1973.
%
\bibitem{bkk} J.\ Binnewies, B.A.\ Kniehl, and G.\ Kramer, {\it Phys. Rev.}
{\bf D52} (1995) 4947.
%
\bibitem{nlounp} P.\ Aurenche, A.\ Douiri, R.\ Baier, M.\ Fontannaz, 
and D.\ Schiff, {\it Phys. Lett.} {\bf 135B} (1984) 164; {\it Nucl. Phys.} 
{\bf B286} (1987) 553; L.E.\ Gordon, {\it Phys. Rev.} {\bf D50} (1994) 6753; 
F.\ Aversa, P.\ Chiappetta, M.\ Greco, and J.Ph.\ Guillet,
{\it Phys. Lett.} {\bf B210} (1988) 225; {\bf B211} (1988) 465;
{\it Nucl. Phys.} {\bf B327} (1989) 105. 
%
\bibitem{dsv} D.\ de\ Florian, M.\ Stratmann, and W.\ Vogelsang,
{\tt hep-ph/9710410}, these proceedings.

\bibitem{jeff} J.R.\ Forshaw and R.G.\ Roberts, {\it Phys. Lett.} {\bf B319},
(1993) 539.
%
\bibitem{PYTHIA} T.\ Sj\"ostrand, {\it Comp.\ Phys.\ Comm.} {\bf 82} 
(1994) 74.
%
\bibitem{safe} W.T.\ Giele and W.B.\ Kilgore, {\it Phys. Rev.} {\bf D55} 
(1997) 7183; W.B.\ Kilgore, talk presented at the `32nd Rencontres de
Moriond: QCD and High-Energy Hadronic Interactions', 1997, Les Arcs,
France, {\tt hep-ph/9705384}; M.H.\ Seymour, {\tt hep-ph/9707338}.
%
\bibitem{Cat93} S.\ Catani, Yu.L.\ Dokshitzer, M.H.\ Seymour, and
B.R.\ Webber, {\it Nucl. Phys.} {\bf B406} (1993) 187.
%
\bibitem{Ell93} S.D.\ Ellis and D.E.\ Soper, 
{\it Phys. Rev.} {\bf D48} (1993)3160.
%
\bibitem{kk} M.\ Klasen and G.\ Kramer, {\it Z. Phys.} {\bf C76} (1997) 67;
B.W.\ Harris and J.F.\ Owens, {\it Phys. Rev.} {\bf{D56}} (1997) 4007.
%
\end{thebibliography}
\end{document}